\documentclass[preprint,showpacs,preprintnumbers,amsmath,amssymb]{revtex4}
\begin{document}

\title{Finite-size and asymptotic behaviors of  
the gyration radius of knotted cylindrical self-avoiding polygons }

\author{Miyuki K. Shimamura}
\email{smiyuki@exp.t.u-tokyo.ac.jp}
\address{
Department of Applied Physics, Graduate School of Engineering,
University of Tokyo, 7-3-1 Hongo, Bunkyo-ku, Tokyo 113-8656, Japan 
}
\author{Tetsuo Deguchi}
\address{
Department of Physics, Faculty of Science, Ochanomizu University,
2-1-1 Ohtsuka, Bunkyo-ku, Tokyo 112-8610, Japan 
}

\begin{abstract}
Several nontrivial properties are shown for  
 the mean square radius of gyration $R_K^2$ 
of ring polymers with a fixed knot type $K$.  
 Through computer simulation, 
we discuss both finite-size and asymptotic behaviors 
 of the gyration radius under the topological constraint 
  for self-avoiding polygons consisting 
 of $N$ cylindrical segments with radius $r$.  
 We find that the average size of ring polymers 
 with a knot $K$ can be much larger than that of 
 no topological constraint.  
The effective expansion due to the topological constraint 
 depends strongly on the parameter $r$ 
which is related to the excluded volume.  
The topological expansion is particularly  
significant for the small $r$ case, 
where the simulation result is associated with 
that of random polygons with the knot $K$. 
\end{abstract}


\maketitle

\section{Introduction}
A  ring polymer is one of the simplest systems 
that have the effect of topological entanglement. 
The  topological state of a ring polymer is given by a knot, 
and it is fixed after the ring polymer is formed. 
The entropy of the ring polymer with the fixed knot 
is much smaller than that of no topological constraint.   
Thus, there should be several nontrivial  properties  
in statistical mechanics of ring polymers with a fixed  topology.  
Furthermore, some dynamical or thermodynamical properties 
of ring polymers under topological constraints 
could also be nontrivial. In fact,  
various computer simulations of ring polymers 
with fixed topology  were performed 
by several groups 
\cite{Vologodskii1,desCloizeaux,LeBret,Chen,Michels,Klenin,Janse1991,Koniaris,DeguchiJKTR,DeguchiRevE,Orlandini,Lai}.
However, there are still many unsolved problems 
related to the topological effect, 
such as the average size of a knotted ring polymer in solution. 

\par 
In the paper, we discuss how the excluded volume controls 
the topological effect on 
 the average size of ring polymers in good solution. 
As a model of ring polymers we employ  
a model of  self-avoiding polygons 
consisting of cylindrical segments with radius $r$.  
 Through numerical simulation,  we investigate 
 the mean square radius of gyration 
 of cylindrical self-avoiding polygons with radius $r$ \cite{PLA,JPSJ,PRE}.  
By changing the cylinder radius $r$, 
we modify the excluded volume effect.  
Thus, we can investigate the topological effect
systematically through 
the simulation of  cylindrical self-avoiding polygons 
for various values of cylinder radius $r$.

\par Let us consider the two cases when 
the radius  $r$ is very large or very small. 
When the radius $r$ is very large, the simulation  
should be related to that of the self-avoiding polygons on the lattice 
\cite{Orlandini,PRE}.  On the other hand, when the radius $r$ is very small, 
 it is related to random polygons with a fixed topology, 
 as we shall see explicitly through the data.  
 In fact, there is quite an interesting suggestion 
\cite{desCloizeaux-Let,Deutsch,Grosberg} that 
under a topological constraint  
the average size of ring polymers with no excluded volume 
should be similar as that of ring polymers with the excluded volume,  
since nontrivial entropic repulsion should be derived  
from the topological constraint. 
According to the suggestion,  the average size 
of random polygons with the trivial knot 
should  be given by  $N^{\nu_{\rm SAW}}$ with respect to the number $N$ 
of polygonal nodes, where $\nu_{\rm SAW}$ is 
the exponent of self-avoiding walks. 
Thus,  the small $r$ case of the simulation in the paper    
should  be important also in the study of  
the topological effect on random polygons.

\par 
The outline of the paper is given in the following. 
In Sec. II we explain self-avoiding polygons (SAPs) 
consisting of cylinder segments.  We also discuss   
 the effective exponent of the mean square radius of gyration 
under no topological constraint $R^2$. 
In Sec. III, we discuss various nontrivial 
finite-size properties of the mean square radius of gyration $R_K^2$ 
for cylindrical SAPs with a given knot type $K$. 
The ratio $R_{K}^2/R^2$  expresses   
the effective expansion due to the topological constraint. 
Through the simulation, we find that the topological effect is important 
 particularly  in  the small $r$ case for cylindrical SAPs.  
Furthermore, the effective topological expansion 
is controlled by the parameter $r$.  
In Sec. IV, we discuss the asymptotic expansion of the ratio 
 $R_{K}^2/R^2$ with respect to the number $N$. 
Finally, in Sec. V, we graphically explain the effective expansion 
of the cylindrical SAPs under the topological constraint, 
through the graphs in the $N-r$ plane.

%
\section{Cylindrical self-avoiding polygons}

\subsection{Cylindrical ring-dimerization algorithm and random knots}

\par 
Let us introduce a model of ring polymers in good solution. 
We consider self-avoiding polygons consisting of 
$N$ rigid impenetrable cylinders of unit length and diameter $r$:   
there is no overlap allowed for any non-adjacent pairs of cylindrical segments, 
while next-neighboring cylinders may overlap each other.  
We call them  cylindrical self-avoiding polygons 
or cylindrical SAPs, for short.  The cylinder radius $r$  
can be related to the stiffness of some stiff polymers such as 
DNAs \cite{Klenin,JPSJ}.   

\par 
In the simulations of the paper, 
we have constructed a large number of cylindrical SAPs 
by the cylindrical ring-dimerization method \cite{PLA}. 
The method is based on the algorithm of ring-dimerization \cite{Chen},  
and very useful for generating long self-avoiding polygons 
(for details, see Ref. \cite{JPSJ}). 
 Here we note that another algorithm 
is discussed in Ref. \cite{Klenin} for the model of cylindrical SAPs, where 
 self-avoiding polygons of impenetrable cylinders with 
 $N < 100$  are constructed  
 in association with knotted DNAs \cite{Rybenkov,Shaw}.

 \par 
In the cylindrical ring-dimerization method, 
 a statistical weight is given to 
 any self-avoiding polygon successfully concatenated.  
Thus, when we evaluate some quantity, 
we take the weighted average of it with respect 
to the statistical weight. Some details 
on the statistical weight of successful concatenation is 
 given in Ref. \cite{JPSJ}. Hereafter in the paper, 
  however, we do not express the statistical weight, for simplicity.

\par 
Let us describe the processes of our  numerical experiments.  
First, we construct $M$ samples of cylindrical SAPs with 
$N$ nodes by the cylindrical ring-dimerization method. 
We put $M=10^4$. Here we note that various knot types are 
included  in the $M$ random samples. 
Second, we make knot diagrams 
for the three-dimensional configurations 
of cylindrical SAPs, by projecting them onto a plain.   
 Then, we calculate two knot invariants $\Delta_K(-1)$ and $v_2(K)$ 
 for the knot diagrams. Third, we select only such polygons  
that have the same set of values of the two knot invariants, 
and then evaluate physical quantities 
such as mean-squared gyration radius for 
the selected cylindrical SAPs.

\par 
The symbol  $\Delta_K(-1)$ denotes the determinant of a knot $K$, 
which is given by the Alexander polynomial $\Delta(t)$ 
evaluated at $t=-1$.  The symbol $v_2(K)$ 
is the  Vassiliev invariant of the second degree \cite{DeguchiPLA,Polyak}.  
The  two knot invariants are 
practically useful for computer simulation 
of random polygons with a large number of polygonal nodes. 
In fact,  it has been  demonstrated in Ref. \cite{DeguchiPLA} 
that the Vassiliev invariant $v_2(K)$ can be calculated 
not only in polynomial time but also without using large memory area.

\subsection{Characteristic length of random knotting $N_c(r)$}

 For a given knot $K$, we consider the probability $P_K(N,r)$ that 
 the topology of an $N$-noded self-avoiding polygon 
with cylinder radius $r$   is given by the knot type $K$. 
We call it the knotting probability of the knot $K$. 
Let us assume that we have $M_K$ self-avoiding polygons 
with a given knot type $K$ among  $M$ samples of cylindrical SAPs 
with radius $r$. Then, we  evaluate 
the knotting probability $P_K(N,r)$ by  
$P_K(N,r) = M_K / M$.

\par  
For the trivial knot,  the knotting probability 
$P_{triv}(N,r)$ for the cylindrical SAPs is given by 
\begin{equation} 
P_{triv}(N,r) = C_{triv} \, \exp(-N/N_c(r)) \, . 
\end{equation} 
Here the estimate of  the constant $C_{triv}$ 
is close to 1.0 \cite{PLA}. 
We call $N_c(r)$ the characteristic length of random knotting. 
It is also shown in Ref. \cite{PLA} that $N_c(r)$  
can be approximated by an exponential function of $r$:
\begin{equation} 
N_c=N_c(0) \exp(\gamma r).
\label{ncr} 
\end{equation}
The best estimates of the two parameters $N_c(0)$ and $\gamma$ 
are given by $N_c(0)=292\pm5$ and 
$\gamma=43.5\pm0.6$  \cite{PLA}.

\par 
For several knots, it is shown \cite{JPSJ} that the knotting probability  
$P_{K}(N,r)$ of a  knot $K$ is given by 
\begin{equation}
P_{K}(N,r)=C_K \left( {\frac N {N_K(r)}} \right)^{m(K)} \exp(-N/N_K(r)) \, . 
\end{equation}
It is numerically suggested in Ref. \cite{JPSJ} 
that $N_K(r)$ should be independent of $K$: 
$N_K(r) \approx N_c(r)$, and also that 
the constant $C_K$ should be independent of the cylinder radius $r$.

\subsection{Mean-squared gyration radius with a topological constraint}
 
The mean square radius of gyration $R^2$ 
of a self-avoiding polygon is defined by 
\begin{equation}
R^2=\frac{1}{2N^2}\sum_{n,m=1}^{N} <( \vec{R}_n-\vec{R}_m)^2> \, . 
\end{equation}
Here $\vec{R}_n$ is the position vector of the $n$th segment 
(or the $n$th node) and $<\cdot>$ denotes the ensemble average, 
which is taken over all possible configurations  of the self-avoiding polygon.

\par Suppose that we have  $M$ self-avoiding polygons. 
Then, we evaluate  the mean square radius of gyration $R^2$  
by the sum: $R^2 =  \sum_{i=1}^{M} R_i^2/M$, 
where $R_i^2$ denotes the gyration radius  
of the $i$th  SAPs in the given $M$ SAPs.

\par 
Let us define the mean square radius of gyration 
$R_K^2$ for such self-avoiding polygons that have a given knot type $K$: 
\begin{equation}
R_K^2=\frac{1}{M_K}\sum_{i=1}^{M_K} R_{K,i}^2,
\end{equation}
where $R_{K,i}^2$ denotes the gyration radius 
of the $i$th self-avoiding polygon that has the knot type $K$.
In terms of $R^2_K$, $R^2$ is given by $R^2=\sum_{K}  R_K^2 M_K/M$.

\par 
 In Fig. 1, the estimates of the mean square radius of gyration $R^2$ 
 are plotted against the number $N$ of nodes in a double-logarithmic scale, 
 for the cylindrical SAPs with $r=0.003$ and $r=0.03$.  
We may confirm the standard asymptotic behaviors 
of the mean-squared gyration radius  $R^2$ in Fig. 1. 
Here we remark on an effective exponent $\nu_{\rm eff}$,  
which is defined through the power-law approximation: $R \sim N^{\nu_{\rm eff}}$.  
It is shown in  \cite{PhD-Miyuki} that  the estimate of 
 the effective exponent $\nu_{\rm eff}$ for the cylindrical SAPs with 
radius $r$ is consistent with  that of the cylindrical SAWs with radius $r$.

\section{Finite-size behaviors of $R_{k}^2$ for some knots }

Let us discuss  simulation results 
on the mean square radius of gyration $R_K^2$   
for the cylindrical SAPs with a knot $K$ and of radius $r$. 
 For two prime knots (the trivial and trefoil knots) 
and a composite knot (the double-trefoil knot,  $3_1 \sharp 3_1$),  
 we have investigated the mean-squared gyration radius $R_K^2$  
under the topological constraint 
in the range of the number $N$ satisfying $21\le N \le 1001$, 
  and for 14 different values of  cylinder radius $r$.

\par 
The gyration radius $R_K^2$ can approximately given by  some power of $N$.  
In Fig. 2, double-logarithmic plots of $R_K^2$ versus  $N$ 
are given for the trivial and trefoil knots, with  two values of 
cylinder radius: $r=0.003$ and 0.03.  
We see that all the double-logarithmic plots of Fig. 2   
fit to some straight lines.  We note that for 
 other values of cylinder radius $r$, 
 several double-logarithmic plots of $R_K^2$ versus  $N$ 
 are explicitly shown in Ref. \cite{PhD-Miyuki}. 

\par 
With the number $N$  fixed, $R_{K}^2$ should increase 
with respect to the radius $r$ for any knot. 
 In Fig. 2, closed squares for $r=0.03$ are located higher 
in the vertical direction than closed circles for $r=0.003$, 
 through  the whole range of $N$ 
both for the trivial and trefoil knots.

\subsection{Ratio $R_K^2/R^2$ and 
the effective  expansion under the topological constraint}

\par 
Let us now consider the ratio of  $R_K^2$ to  $R^2$ 
for a given knot $K$.  
If the ratio is larger (smaller) than 1.0, then the average size 
of SAPs with the knot $K$ is relatively larger (smaller) than that of 
no topological constraint.  We say that 
the SAPs with the knot $K$ is  effectively 
more (less) expanded. In Fig. 3,  
the ratio $R_K^2/R^2$ versus the number $N$ is plotted 
in a double-logarithmic scale  
for the trivial and trefoil knots. 
Here, we have depicted only the case of $r=0.003$ 
among many sets of the cylindrical SAPs with the 14 different values of 
 cylindrical radius. 

\par 
 For the trivial knot, we see in Fig. 3 that 
the ratio $R_{triv}^2/R^2$ is greater than 1.0 when $N > 50 $. 
Thus, the average size of the ring polymers  with the trivial knot  
enlarges under the topological constraint. 
 It gives a typical example of effective expansion.

\par 
 In Fig. 3,  the graph of the trivial knot 
is  convex downwards: 
the ratio $R_{triv}^2/R^2$ is almost constant with respect to $N$ 
for small $N$ such as $N < 100$; for $N > 300$ 
 the ratio $R^2_{triv}/R^2$ increases with respect to $N$  
 with a larger gradient,  and 
the graph can be approximated  by a power law  
 such as $R_{triv}^2/R^2 \propto N^{\nu_{\rm eff}^{triv}}$. 
Here the symbol  $\nu_{\rm eff}^{triv}$ 
 denotes the effective exponent for the trivial knot. 
We note that the characteristic length $N_c(r)$ is 
 approximately given  by 300 for $r=0.003$. 
Thus, we may say that the power law behavior is valid 
for  $N > N_c(r)$.

\par 
 For the trefoil knot, the graph can be approximated 
 by a power of $N$ 
 such as $R_{tre}^2  \propto  N^{\nu_{\rm eff}^{tre}}$ through 
 the range of $100 \le N \le 1001$.   
Here the symbol $\nu_{\rm eff}^{tre}$ denotes the effective exponent 
of the trefoil knot.  
 In Fig. 3, we  find that 
 when $N < 100$ the ratio $R_{tre}^2/R^2$ is smaller than 1.0, while 
it is larger than 1.0 when $N > 300$. 
Thus, when $N$ is small,  the topological constraint of the trefoil knot 
gives effective shrinking to ring polymers,    
while it does not when  $N$ is large. 
 For a nontrivial knot $K$, we expect that  
 the ratio $R^2_{K}/R^2$  is less than 1.0  
 when $N$ is small,  while it can be  larger than 1.0 when  $N$ is large.

\par 
The properties of the ratio $R^2_K/R^2$ discussed in the last three paragraphs 
 are  consistent with the  
 simulation results of Gaussian random polygons \cite{Miyuki}. 
We have found for the random polygons  that 
the double-logarithmic graph of $R_{K}^2/R^2$ versus $N$ 
is given by a downward convex curve for the trivial knot, 
while it is given by a straight line for the trefoil knot 
 and also for  other several nontrivial knots;     
 for the nontrivial knots investigated, the ratio 
 $R_{K}^2/R^2$ is given by some power of $N$ 
 such as $N^{\nu_{\rm eff}^{K}}$ .  
Thus, there are indeed many important properties 
valid both for the simulation of the Gaussian random polygons 
and that of the cylindrical SAPs with a small radius such as $r=0.003$. 

\par 
The observations derived from Fig. 3 
should be  valid particularly for finite-size systems.     
Admitting that $N$ is finite, 
we can only understand that  the Gaussian random polygons 
and the cylindrical SAPs have the similar topological properties 
in common. If we discuss asymptotic 
behaviors, SAPs and random polygons should be quite different.  
However, if we consider such properties that are valid 
for finite $N$, then they can hold  
both for SAPs with small excluded volume 
and random polygons that have no excluded volume.

\par 
Let us discuss again the convexity of the graph of the trivial knot, 
which has been observed in Fig. 3. 
We consider how the convexity depends on the radius $r$.  
In Fig. 4,  the graphs of the ratio $R_{triv}^2/R^2$ versus $N$ 
are given in a double-logarithmic scale 
for four different values of cylinder radius $r$.   
Then, we see that the graph with $r=0.05$ 
is less convex than that of $r=0.003$. 
Thus,  the convexity in the graphs of the effective expansion 
for the trivial knot 
should  be valid only when cylinder radius  $r$ is small.

\par 
Let us assume  that the convexity of the graphs of
 $R_{triv}^2/R^2$  for the small $r$ case 
should correspond to  a crossover behavior 
of $R_{triv}^2/R^2$ with respect to $N$. 
Then, the crossover behavior could  be related to 
that of Gaussian random knots, which is  recently discussed by  
Grosberg \cite{Grosberg} for Gaussian random polygons. 
We can discuss the convexity of the double-logarithmic 
graph of $R_{triv}^2/R^2$ versus $N$,  
taking an analogy with the crossover 
of the Gaussian random knots. 
Thus,  we call the convexity  of the trivial knot in Fig. 3 
the crossover, hereafter in the paper.

\par 
 For the non-trivial knots investigated, we do not see 
any crossover in the graph of $R_{K}^2/R^2$ versus $N$. 
 For instance, for the $4_1$ and $3_1 \sharp 3_1$ knots, 
 the slope of the graph near $N \sim N_c(r)$ is straight 
 in the double-logarithmic scale. 
The  crossover at $N \sim N_c(r)$ should be 
valid only for the trivial knot.

\subsection{The plateau in the graph of $R_K^2/R^2$ versus $N$ for large $N$}

\par  
We discuss how the ratio $R_K^2/R^2$ depends on the number $N$,     
considering both the excluded volume effect and 
the effective expansion due to the topological constraint. 
In Fig. 5, the graphs of the ratio $R_{K}^2/R^2$  versus $N$ 
for different values of cylinder radius $r$ 
 are shown in linear scales: (a) for 
  the trivial knot; (b) for the trefoil knot.

\par 
Let us first consider the large $N$ behaviors  of the graphs 
 shown in Fig. 5 for the trivial and trefoil knots. 
The graphs of $R_K^2/R^2$ versus $N$ 
have a common tendency that they become constant with respect 
to $N$ when $N$ is very large. 
It is particularly the case for the larger values of 
cylinder radius $r$ such as $r=0.03$ and 0.05.   
They approach  horizontal lines at some large values of $N$. 
When $r$ is small such as  $r=0.003$,
the graph becomes flat only for large $N$, 
as shown in  Fig. 5.

\par 
  From the flatness of the graphs of $R_K^2/R^2$ for large $N$, 
 it follows that the power law behavior: 
 $R_{tre}^2/R^2 \propto N^{\nu_{\rm eff}^{tre}}$ 
does not hold when $N$ is very large. 
In Fig. 3, we have discussed  that 
the ratio $R_{tre}^2/R^2$ versus the number $N$ 
can  be  approximated by the power law  for $r=0.003$ 
through the range of $100 \le N \le 1001$. 
However, the power-law approximation should be valid 
only within some finite range of $N$.

\par 
 Let us discuss other finite-$N$ behaviors of the ratio $R_K^2/R^2$. 
 For the trefoil knot, the ratio $R_{tre}^2/R^2$ is less than 1.0 when $N$ is small; 
it approaches or becomes larger than 1.0 when $N$ is large enough.  
 When cylinder radius $r$ is small such as $r =0.003$ and $r=0.01$,  
the ratio $R_{tre}^2/R^2$ is clearly greater than 1.0 when $N$ is large enough.    
When $r$ is small,  there should be a critical value $N_{critical}$ such that 
 $R_{tre}^2/R^2 < 1.0$ for $N < N_{critical}$, and  
 $R_{tre}^2/R^2 > 1.0$ for $N > N_{critical}$. 
 Furthermore, we have a conjecture that the critical value 
$N_{critical}$ should be roughly equal  to 
 the characteristic length $N_c(r)$ of random knotting. 
It seems that the conjecture is consistent with the graphs of Fig. 5 (b).    

\par 
Let us discuss the conjecture on $N_{critical}$, explicitly. 
In Fig. 5 (b),  we see that  
 for $r=0.003$, the ratio  $R_{tre}^2/R^2$ 
 becomes 1.0 roughly at $N=300$, and also that   
 for $r=0.01$, the ratio  $R_{tre}^2/R^2$ is close to 1.0 roughly at $N=400$. 
The observations are  consistent with the estimates 
of $N_c(r)$ in Ref. \cite{PLA}: 
$N_c(r) = (2.72 \pm 0.06 )\times 10^2$ for $r=0.0$ and  
$N_c(r) = (4.72 \pm 0.14) \times 10^2$ for $r=0.01$.  
Thus, the consistency supports the conjecture on $N_{critical}$.

\subsection{Decrease of the topological effect  
under the increase of the excluded volume }

\par  
The effect of a topological constraint on the gyration radius  decreases   
 when the excluded volume increases. There are two examples:  
 the decrease of ratio $R_K^2/R^2$ with respect to 
cylinder radius $r$ while $N$ being fixed,  and  
the disappearance of the crossover for the trivial knot shown in 
Figs. 3 and 4.

\par 
Let us first discuss  how the excluded-volume can modify  
  the effective expansion  due to the topological constraint.    
As we clearly see in Fig. 5,  
the ratio  $R_{K}^2/R^2$  
decreases as cylinder radius $r$ increases with  $N$ fixed, 
both for the trivial and trefoil knots. 
Thus, the effective expansion of SAPs 
under the topological constraint 
becomes smaller  when  the excluded volume becomes larger.

\par 
It is quite nontrivial that 
the effective expansion given by the ratio $R_K^2/R^2$
decreases as cylinder radius $r$ increases.   
In fact, the value of $R_{K}^2$ 
 itself increases with respect to $r$, as we have observed in Fig. 2.  
Furthermore, one might expect that the effective expansion 
due to a topological constraint should  also increase 
with respect to cylinder radius $r$, simply  because 
the average size of ring polymers with larger excluded volume  
becomes larger, as observed in Fig. 1.  
However, it is not the case for the ratio $R_K^2/R^2$.

\par 
Let us now discuss the crossover behavior of the trivial knot again, 
from the viewpoint of the competition between the topological effect 
and the excluded volume effect.  Here we recall that the crossover 
has been discussed in \S 3.A with Figs. 3 and 4.    
Here we regard the crossover as a characteristic behavior derived from 
the topological constraint of being the trivial knot.  

\par 
As a working hypothesis, let us assume that 
 the crossover should occur at around the characteristic length $N_c(r)$.  
Recall that $N_c(r)$ is larger than 1000 for $r=0.03$ and 0.05, 
as we have estimated: $N_c(r) \approx 1200$ for $r=0.03$, and  
$N_c(r) \approx 2600$ for $r=0.05$. 
If the above hypothesis would be valid, 
then the graphs for $r$ = 0.03 and 0.05 
 should also be convex. 
In Fig. 4,  however,  we see  no change 
in the gradient of the graph 
of  $R_{triv}^2/R^2$ versus $N$ for $r=0.03$ or 0.05. 
The assumed crossover of the trivial knot does not 
appear for $r$ = 0.03 or 0.05.   
We may thus consider  
that the crossover as a topological effect 
is  diminished by the excluded volume effect 
when $r \ge 0.03$.

\subsection{Characteristic length of random knotting $N_c(r)$ and the effective expansion}

In terms of the characteristic length $N_c(r)$, we can  
explain some  properties of 
the effective expansion of cylindrical SAPs 
under a topological constraint.  
Here we recall that the ratio $R_{K}^2/R^2$ describes the degree of 
the effective expansion under the topological constraint of a knot $K$.  

\par 
We first consider the case when the characteristic length  
$N_c(r)$ is very large.  Let us show that 
 the ratio $R_{triv}^2/R^2$ should be close to 1.0 for $N \ll N_c(r)$.  
 First, we recall that 
the probability $P_{triv}(N)$ of the trivial knot decays  
 exponentially  with respect to the  
number $N$ of polygonal nodes: 
$P_{triv}(N) = \exp(-N/N_c(r))$. If $N/N_c(r)$ is very small, 
the probability  $P_{triv}(N)$ is close to 1.0, i.e.,  
almost all SAPs have the trivial knot.    
Then, the mean-squared gyration radius  
with no topological constraint $R^2$ 
should be almost equal to that of the trivial knot $R_{triv}^2$.  
 Consequently, the ratio $R_{triv}^2/R^2$ should  be close to 1.0.

\par 
 When $r \ge 0.05$, the characteristic length $N_c(r)$ 
is larger than $2600$.  
Then,  the trivial knot is dominant among the possible knots generated  
in SAPs with $N <1000$. 
Thus, $R_{triv}^2$ should almost agree with $R^2$, which is 
the mean-squared gyration radius of SAPs under no topological constraint.  
There is  no effective expansion under the topological constraint: 
the $R_{triv}^2/R^2$ is close to 1.0.

\par 
Let us next consider the case when the characteristic length  
$N_c(r)$ is small or not large. Then we show that 
the mean square radius of gyration 
of SAPs with the trivial knot  $R_{triv}^2$  should be  
larger than that of no topological constraint $R^2$ 
for $N > N_c(r)$. 
In fact,  various types of knots can 
appear in a given set of randomly generated SAPs of the cylinder radius $r$, 
since  the probability of the trivial knot $P_{triv}(N)$ 
is exponentially  small for $N > N_c(r)$ . 
We note that the fraction of 
nontrivial knots is given by $1- \exp(-N/N_c(r))$.  
Thus, it is not certain whether the ratio $R_{triv}^2/R^2$ is 
close to the value 1.0 or not. However, 
we may expect that the ratio 
$R_{triv}^2/R^2$ should be indeed larger than 1.0.   
Here we  consider the following points: 
 when $N > N_c(r)$, the majority of  SAPs generated randomly 
should  have much more complex knots than the trivial knot; 
 the mean square radius of gyration 
of  $N$-noded SAPs with a very complex knot 
should be much smaller than that of the trivial knot.

\par 
The  explanation on the effective expansion  
discussed in the above  is completely 
consistent with the simulation results,  
as having been discussed in \S 3, in particular,  
through Figs. 3, 4 and 5.

%
%
\section{Asymptotic behaviors of $R_K^2$ }
\subsection{The  exponent of $R^2_{K}$ }

Let us discuss an asymptotic expansion for the mean square radius of gyration  
of cylindrical SAPs with  a given knot $K$. 
Here we assume that $R_K^2$  can be expanded 
in terms of $1/N$ consistently 
with renormalization group arguments. 
Then,  the large $N$ dependence of $R_K^2$ is given by 
\begin{equation} 
R_K^2=A_K N^{2\nu_K} \Bigl[ 1+B_K N^{-\Delta} +O(1/N) \Bigr].
\label{asymptotic} 
\end{equation}
Here, the exponent $\nu_K$ should be given by that of self-avoiding walks:
$\nu_K = \nu_{\rm SAW}$. 
In order to analyze the numerical data systematically,  
however, we have introduced $\nu_K$ as a fitting parameter.   
Thus, for the ratio $R_K^2/R^2$,   
we have the following expansion:
\begin{equation}
\label{rg}
R_K^2/R^2=(A_K/A) N^{2\Delta \nu_K} 
\Bigl[ 1+(B_K-B) N^{-\Delta} +O(1/N) \Bigr] \, . 
\end{equation}
Here we have put $\Delta \nu_K$ as a fitting parameter.

\par 
We have analyzed the data for the three different knots: 
the trivial, trefoil and $3_1 \sharp 3_1$ knots,  
applying the expansion (\ref{rg})   
to the numerical data of $R_{K}^2/R^2$ for $N \ge 300$ . 
The best estimates of the three parameters  
are given in Tables 1, 2 and 3 for the trivial, trefoil and 
$3_1 \sharp 3_1$ knots, respectively. 

\par 
Let us discuss the best estimates of the difference of the exponents: 
$\Delta \nu_K$. 
We see in Tables 1 , 2 and 3 that all the results of $\Delta \nu_K$ suggest 
that they should be given by 0.0, with respect to the confidence interval. 
Let us examine the best estimates more precisely. 
It is rather clear from Tables 1, 2 and 3 that  for a given cylinder radius $r$,  
the best estimates of  $\Delta \nu_K$ 
 are independent of the knot type.  

\par 
 There is  another evidence supporting that $\Delta \nu_K=0.0$ for the trivial and trefoil knots.
Let us consider the plots of the ratio $R_K^2/R^2$ versus $N$ in Fig. 5 
for the trivial and trefoil knots. 
We recall that the graphs are likely to approach some horizontal lines at some large $N$.
The tendency of the graphs becoming flat for large $N$ 
suggests that  $R_K^2$ and $R^2$ should have the same exponent, 
i.e., $\nu_{\rm SAW}$.   

\par 
 From the two observations, we  conclude that 
the difference of the exponents is given by 0.0: $\Delta \nu_K = 0.0$ 
for any value of $r$.     
There is thus no topological effect on the scaling exponent 
defined in the asymptotic expansion of $R^2_K$.

\subsection{Amplitude ratio $A_K/A$ }

\par 
Let us now consider  the amplitude $A_K$ of the asymptotic expansion (\ref{rg}). 
 In Tables 1, 2 and 3,  the best estimates of  
 the ratio  $A_K/A$ are larger than 1.0 for the three knots, when $r$ is small. 
 The observation must be  important. In fact,   
 if the amplitude ratio $A_K/A$ is larger than 1.0 in the asymptotic expansion 
 (\ref{rg}), then $R^2_K$ is larger than $R^2$ for any large value of $N$. 
 It might seem that the consequence  is against the standard thermodynamic limit 
 of statistical mechanics.

\par 
 However, there is a clear evidence for 
 the observation that $A_K/A > 1.0 $ for some small values 
 of cylinder radius $r$. 
  In fact,  the graphs of the ratio $R_{K}^2/R^2$ versus $N$ 
 are monotonically increasing with respect to $N$, as we see in Figs. 3, 4 and 5.  
 It is clear that the graphs with the smaller values of cylinder radius $r$ 
 are larger than 1.0 when  $N$ is large. 
 This observations of Figs. 3, 4 and 5 confirm that  $A_K/A > 1.0 $ 
 when cylinder radius $r$ is small. 
 Thus, we may conclude that  the topological constraint 
 gives an effective expansion also to asymptotically large cylindrical SAPs 
  when the radius $r$ is small.

\par 
The value of  $A_K/A$  decreases with respect to the radius $r$ 
for the three knots.  We see it 
in Tables 1 to 3, where the best estimates of $A_K/A$ are listed. 
It is also consistent with the fact that 
the ratio $R_{K}^2/R^2$ decreases with respect to $r$, which we have discussed 
 in \S 3.C. 
 However, the decrease of  $A_K/A$  is quite nontrivial, 
  since  the mean-squared gyration radius $R_K^2$ itself increases with respect to  $r$, 
for the trivial and trefoil knots, as shown in Fig. 2. 
Here we recall in Fig. 1 that the gyration radius under no topological constraint $R^2$ increases 
with respect to $r$.

\par 
 From the viewpoint of asymptotic behaviors,  we have shown 
that the effective expansion derived from the topological repulsion 
decreases with respect to cylinder radius $r$. We have also discussed  that 
$R^2_K$ is larger than $R^2$ for any large value of $N$, when 
cylinder radius $r$ is small.

\subsection{The $r$-dependence of the amplitude ratio}

Let us discuss  the $r$ dependence of 
the amplitude ratio $A_K/A$, more quantitatively.  
For this purpose, we analyze the data of $R^2_K/R^2$ versus $N$ again, 
 assuming  $\nu_K=\nu$ in eq. (\ref{rg}). 
We evaluate the amplitude  ratio $A_K/A$  
by the following formula:
\begin{equation} 
R_K^2/R^2=\alpha_K (1+ \beta_K N^{-\Delta}+O(1/N)) \, .
\label{twopt} 
\end{equation} 
Here we have replaced with $\alpha_K$ and $\beta_K$, 
  $A_K/A$ and $B_K-B$ in (\ref{rg}), 
respectively.  Here we have also introduced a technical assumption: 
 $\Delta=\Delta_K$=0.5 in  (\ref{rg}). 

\par 
We have obtained the numerical estimates of  $\alpha_K$,
 applying the fitting formula 
(\ref{twopt}) to  the data of $R_{K}^2/R^2$ with  $N \ge 300$ .
The estimates of  $\alpha_K$ versus $r$ 
are shown in Fig. 6 in the double-logarithmic scale 
for the  trivial, trefoil, and $3_1 \sharp 3_1$ knots. 
To be precise, 
the values of $\alpha_K$ are a little larger 
than those of $A_K/A$ given in Tables 1 , 2 and 3.

\par 
The estimate of the parameter $\alpha_K$  becomes 
close to the value 1.0  when  cylinder radius $r$ 
is large enough. Furthermore, 
it is suggested from Fig. 6 that $\alpha_K$ should be independent 
of the knot type.  In fact, the data points for 
the trivial, trefoil and the double-trefoil ($3_1 \# 3_1$) 
knots overlap each other. 
These two observations 
are consistent with the simulation result of the self-avoiding polygons 
on the  lattice \cite{Orlandini,Janse1991}.   

\par 
Interestingly, we see in Fig. 6 that the ratio $\alpha_K$ 
decreases monotonically with respect to the cylinder radius $r$.
For the data with $0.001 \le r \le 0.01$, 
we find that $\alpha_K$ is roughly approximated 
by a decreasing function of $r$ such as 
$\alpha_K=\alpha_0 r^{\phi} \exp(-\psi r)$, 
with $\alpha_0$=1.00$\pm$0.12, $\phi$=-0.05$\pm$0.02 
and $\psi$=5.78$\pm$4.79.
The $\chi^2$ value is given by 1.

%
%
\section{Discussion}

\par 
With some graphs in the $N-r$ plane, 
we can illustrate the finite-size behaviors of the ratio $R_K^2/R^2$ discussed 
in \S 3.  We recall that  the topological effect 
has  played a central role as well as the excluded-volume effect.  
Thus, we consider  two lengths with respect to 
the number $N$ of polygonal nodes: 
the characteristic length of random knotting  $N_c(r)$ 
and  the ``excluded-volume length'' $N_{ex}(r)$. 
When $N > N_{ex}(r)$,  the excluded-volume effect 
should be important to any $N$-noded SAP with radius $r$ .  

\par 
We define $N_{ex}(r)$ by  $N_{ex}(r) = 1/r^2$. 
The derivation is given in the following.  
We first note that the parameter $z$ of the excluded-volume  is given by 
$ z  =  {\rm Const.} \,  \sqrt{N} B /{\ell^3} 
 \propto  N^{1/2} \, r $,  
where the cylindrical segments have the diameter $d$ and the length $\ell$, 
and the second virial coefficient $B$ of a polymer chain 
is given by $\ell^2 d$ \cite{Grosberg-book}.  
Here we also note that the ratio $d / \ell$ corresponds 
to the radius $r$ of the cylindrical SAPs.   
We may consider that when $z \approx 1$,  
the excluded volume can not be neglected.  Thus we have  
the number $N_{ex}(r)$ 
from the condition:  $\sqrt{ N_{ex}(r)} \, r =1$.

\par 
 We consider two graphical lines  in the $N-r$ plane:  
$N = N_{ex}(r) $ and $N=N_c(r)$. In Fig. 7, the vertical line 
of the diagram expresses 
the $r$-axis and the horizontal one the $N$-axis.   
The graph $N_c(r)=N$ reaches the $N$ axis  
at $N= N_c(0) \approx 300$. 
Here we recall that the function $N_c(r)$ is given by 
eq. (\ref{ncr}): $N_c(r)= N_c(0) \exp\left(\gamma r \right)$. 
There is a  crossing point for the two curved lines. 
The coordinates of the crossing point  
 is approximately given by $N^{*}=1300$ and $r^{*}=0.03$.  
For a given simulation of the ratio $R_K^2/R^2$ with a fixed radius $r$,  
we have a series of   data points located  on a 
straight line parallel to the $N$ axis.

\par 
Let us first consider the case of small values of $r$ 
such as $r=0.003$ and $r=0.01$. From the simulation of \S 3, it is shown that 
the effective expansion due to the topological constraint is large.  
This is consistent with the following interpretation of 
the $N-r$ diagram: 
if we start from the region near the $r$ axis and move in the direction of 
the $N$ axis, then 
we cross the line $N= N_c(r) $ before reaching 
another one  $\sqrt{N} \, r =1$; thus, we expect that  
the excluded-volume  remains small 
when the topological effect becomes significant.

\par The above explanation should be consistent with the observation 
that the crossover of the trivial knot 
occurs  near $N=N_c(r)$ for small values of $r$. Here we recall Figs. 3 and 4. 
When $r$ is very small, then we cross 
the line of $N_c(r) = N $ almost at  $N_c(0) \approx 300$.

\par 
When  radius  $r$ is large  such as $r=0.03$ and 0.05, 
it is shown in \S 3  through simulation 
that the effective expansion is small: 
the ratio $R_{K}^2/R^2$ is close to 1.0.  
In the $N-r$ diagram, when we move rightwards from the region 
near the $r$ axis with $r$ fixed, 
we cross the line $\sqrt{N} \, r =1$ before reaching another  
line  $N_c(r) = N $. Thus, the effective expansion 
as the topological effect should be  small.

\par 
Finally, we should remark that 
some important properties of $R_K^2$ of cylindrical SAPs with radius $r$ 
have been discussed  systematically 
through scaling arguments with the blob picture 
by Grosberg \cite{Grosberg2}. In the note \cite{Grosberg2}, 
the characteristic length $N_c(r)$ and 
the excluded-volume parameter $z$ are explicitly 
discussed in the $N-r$ diagrams. 
It would thus be an interesting future 
problem to investigate how far the predicted properties 
of $R_K^2$  are consistent with simulation results. 

\begin{acknowledgements} 
We would like to thank  Prof. K. Ito  for  helpful discussions.  
We would also like to thank Prof. A.Yu. Grosberg  
for his helpful discussion and sending the unpublished note 
on $R_K^2$ \cite{Grosberg2}.

\end{acknowledgements}


%
%
\newpage 


%
\begin{figure}
\caption{Double-logarithmic plots of the 
mean square radius of gyration under no topological constraint $R^2$ 
for cylindrical SAPs versus the number $N$ of polygonal nodes 
for radius $r=0.003$ and $0.03$ depicted by 
closed circles and squares, respectively. }
\end{figure}

%
\begin{figure}
\caption{Double-logarithmic plot of $R_K^2$ versus $N$ 
 with  $r=0.003$ and $0.03$  
  shown by  closed circles and squares, respectively: 
 (a) for the trivial knot; (b) for the trefoil knot. }
\end{figure}

%
\begin{figure}
\caption{Double-logarithmic plots of 
the ratio $R_K^2/R^2$ versus $N$ for cylindrical SAPs with $r=0.003$. 
$R_{triv}^2/R^2$ and $R_{tre}^2/R^2$ are shown by 
closed circles squares, respectively. }
\end{figure}

%
\begin{figure}
\caption{Double-logarithmic plots of the ratio $R_{triv}^2/R^2$ 
versus  $N$ for $r=$0.003, 0.01, 0.03 and 0.05 
 shown by closed circles, squares,  diamonds and triangles, respectively. }
\end{figure}

%
\begin{figure}
\caption{Graphs of the ratio $R_K^2/R^2$ versus 
the number  $N$ in linear scales 
for $r= 0.003$, 0.01, 0.03 and 0.05 
shown by closed circles, squares, diamonds and triangles, respectively: 
 (a) for the trivial knot and (b) for the trefoil knot. 
The same  data points  are shown in both Fig. 4 and Fig. 5 (a). }
\end{figure}

%
\begin{figure}
\caption{Double-logarithmic plots of the amplitude ratio  
 $\alpha_K$ versus cylinder radius $r$ 
for the trivial, trefoil and  double-trefoil ($3_1 \# 3_1$)  knots shown by 
closed circles, squares and triangles, respectively. 
For the double-trefoil knot, 
the data points for $0.001 \le r \le 0.01$ are shown. }
\end{figure}

%
\begin{figure}
\caption{$N-r$ diagram. Graphs of 
 $N=N_{c}(r)$ and $N=N_{ex}(r)$ are shown by two curved lines. 
  The arrows (a) and (b) suggest 
  the series of the data points of Fig. 5 for $r=0.01$ and $r=0.005$, respectively. 
All the data points in the paper 
are located in the shaded area.  }
\end{figure}

%
%
\newpage 
\begin{table}
\caption{Fitting parameters $A_K/A$, $B_K-B$ and $\Delta \nu_K$ 
versus cylinder radius $r$: for the trivial knot.}
\label{Tab1}
\begin{ruledtabular}
\begin{tabular}{ccccc}
\hline
$r$ & $A_K/A$ & $B_K-B$ & 2$\Delta \nu_K$ & $\chi^2$ \\
\hline
0.001 & 1.313$\pm$1.285 & -2.587$\pm$4.211 & 0.003$\pm$0.123 & 18\\
0.002 & 1.235$\pm$1.152 & -2.389$\pm$4.117 & 0.009$\pm$0.116 & 12\\
0.003 & 1.213$\pm$1.065 & -2.317$\pm$3.912 & 0.009$\pm$0.109 & 3 \\
0.004 & 1.228$\pm$1.062 & -1.982$\pm$3.973 & 0.003$\pm$0.107 & 3 \\
0.005 & 1.170$\pm$0.983 & -1.684$\pm$3.985 & 0.007$\pm$0.104 & 4 \\
0.006 & 1.207$\pm$0.921 & -2.204$\pm$3.464 & 0.005$\pm$0.095 & 4 \\
0.007 & 1.159$\pm$0.891 & -1.633$\pm$3.684 & 0.005$\pm$0.095 & 3 \\
0.01 & 1.106$\pm$0.836 & -1.090$\pm$3.809 & 0.005$\pm$0.092 & 3 \\
0.02 & 1.065$\pm$0.686 & -0.699$\pm$3.384 & 0.003$\pm$0.078 & 1 \\
0.03 & 1.063$\pm$0.628 & -0.518$\pm$3.166 & -0.001$\pm$0.071 & 2 \\
0.04 & 1.043$\pm$0.590 & -0.353$\pm$3.076 & -0.001$\pm$0.068 & 1 \\
0.05 & 1.010$\pm$0.554 & -0.143$\pm$3.039 & 0.002$\pm$0.066 & 1 \\
0.06 & 1.020$\pm$0.551 & -0.103$\pm$2.997 & -0.001$\pm$0.065 & 1  \\
0.07 & 1.013$\pm$0.531 & -0.187$\pm$2.898 & -0.001$\pm$0.060 & 1 \\
\hline
\end{tabular}
\end{ruledtabular}
\end{table}

\begin{table}
\caption{Fitting parameters $A_K/A$, $B_K-B$ and $\Delta \nu_K$ 
versus cylinder radius $r$: for the trefoil knot}
\label{Tab2}
\begin{ruledtabular}
\begin{tabular}{ccccc}
\hline
$r$ & $A_K/A$ & $B_K-B$ & 2$\Delta \nu_K$ & $\chi^2$ \\
\hline
0.001 & 1.286$\pm$0.970 & -4.440$\pm$2.784 & 0.014$\pm$0.096 & 3\\
0.002 & 1.215$\pm$0.918 & -4.093$\pm$2.906 & 0.015$\pm$0.095 & 10 \\
0.003 & 1.202$\pm$0.905 & -3.562$\pm$3.054 & 0.011$\pm$0.094 & 11 \\
0.004 & 1.176$\pm$0.872 & -3.423$\pm$3.069 & 0.012$\pm$0.093 & 16 \\
0.005 & 1.176$\pm$0.845 & -3.461$\pm$2.964 & 0.010$\pm$0.090 & 6 \\
0.006 & 1.113$\pm$0.807 & -3.174$\pm$3.091 & 0.015$\pm$0.090 & 7 \\
0.007 & 1.084$\pm$0.765 & -3.219$\pm$2.996 & 0.019$\pm$0.088 & 3 \\
0.01 & 1.103$\pm$0.743 & -3.220$\pm$2.870  & 0.013$\pm$0.084 & 1 \\
0.02 & 1.068$\pm$0.765 & -2.326$\pm$3.353 & 0.005$\pm$0.088 & 2 \\
0.03 & 1.058$\pm$0.790 & -2.262$\pm$3.531 & 0.003$\pm$0.091 & 4 \\
0.04 & 1.003$\pm$0.835 & -2.043$\pm$4.034 & 0.007$\pm$0.101 & 3 \\
0.05 & 1.007$\pm$0.883 & -2.422$\pm$4.119 & 0.007$\pm$0.107 & 4 \\
0.06 & 1.029$\pm$0.975 & -2.900$\pm$4.274 & 0.005$\pm$0.116 & 3 \\
0.07 & 0.998$\pm$1.197 & -1.923$\pm$5.915 & 0.002$\pm$0.146 & 2 \\
\hline
\end{tabular}
\end{ruledtabular}
\end{table}

\begin{table}
\caption{Fitting parameters $A_K/A$, $B_K-B$ and $\Delta \nu_K$ 
versus cylinder radius $r$:  for the double-trefoil knot ($3_1 \sharp 3_1$).}
\label{Tab3}
\begin{ruledtabular}
\begin{tabular}{ccccc}
\hline
$r$ & $A_K/A$ & $B_K-B$ & 2$\Delta \nu_K$ & $\chi^2$ \\
\hline
0.001 & 1.269$\pm$1.158 & -5.203$\pm$3.255 & 0.012$\pm$0.116 & 7 \\
0.002 & 1.224$\pm$1.077 & -5.203$\pm$3.113 & 0.016$\pm$0.112 & 7 \\
0.003 & 1.158$\pm$1.090 & -4.371$\pm$3.662 & 0.016$\pm$0.118 & 1 \\
0.004 & 1.149$\pm$1.054 & -4.866$\pm$3.401 & 0.018$\pm$0.116 & 9 \\
0.005 & 1.137$\pm$1.008 & -4.851$\pm$3.297 & 0.016$\pm$0.112 & 4 \\
0.006 & 1.096$\pm$1.002 & -4.745$\pm$3.476 & 0.021$\pm$0.115 & 1 \\
0.007 & 1.061$\pm$1.043 & -3.819$\pm$4.091 & 0.020$\pm$0.122 & 3 \\
0.01 & 1.076$\pm$1.369 & -3.563$\pm$5.287 & 0.012$\pm$0.159 & 4\\
\hline
\end{tabular}
\end{ruledtabular}
\end{table}

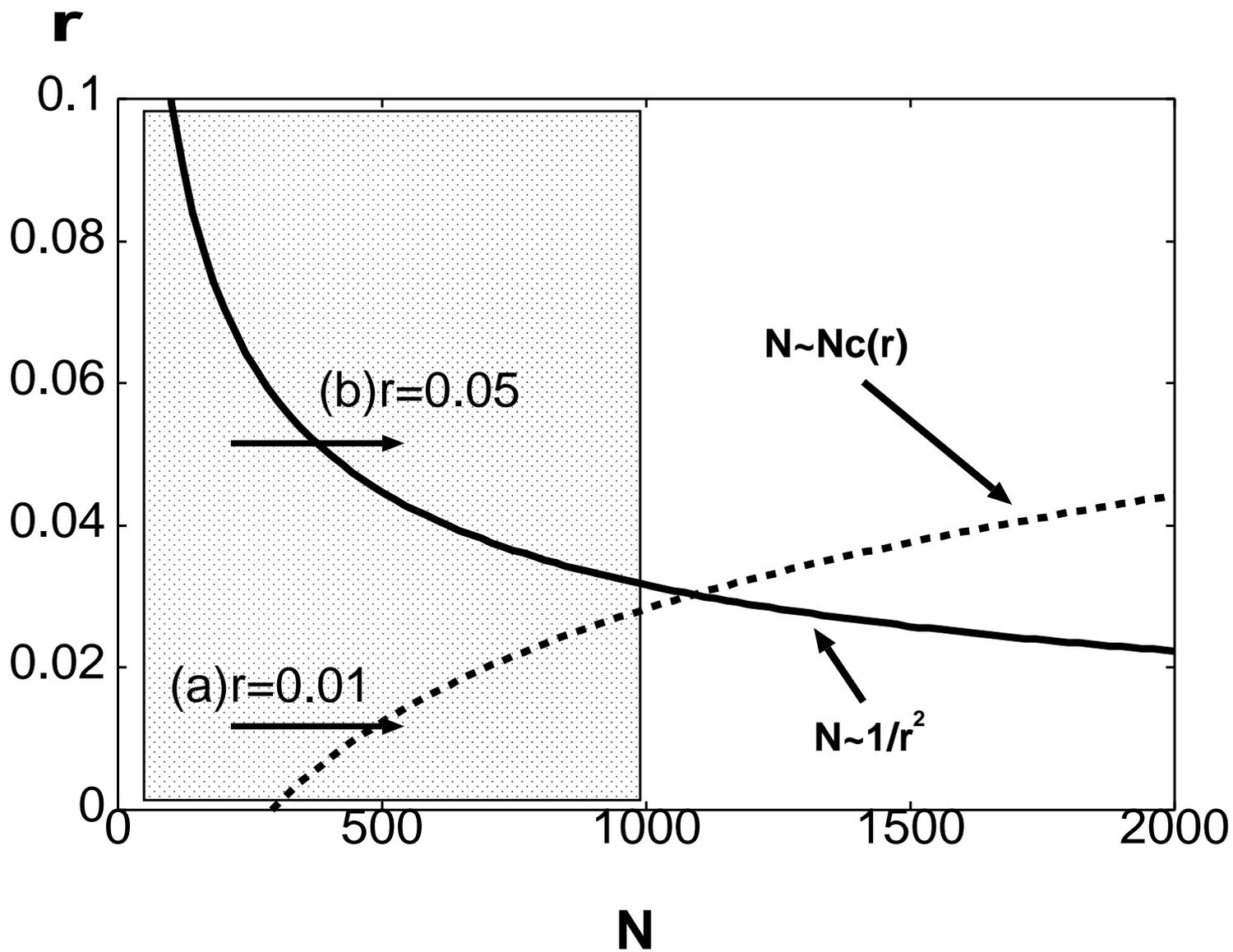
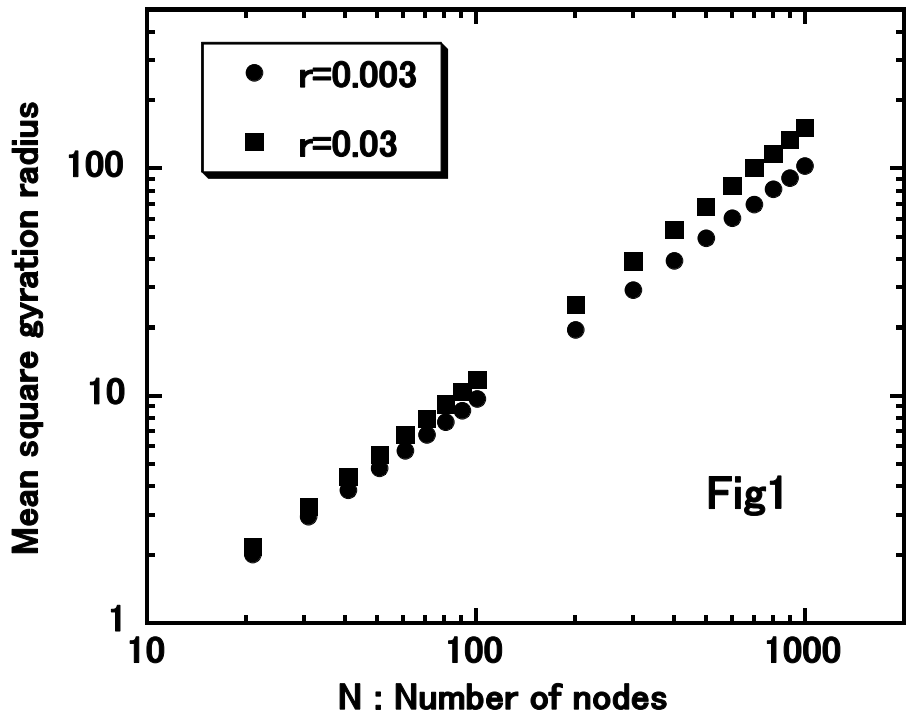
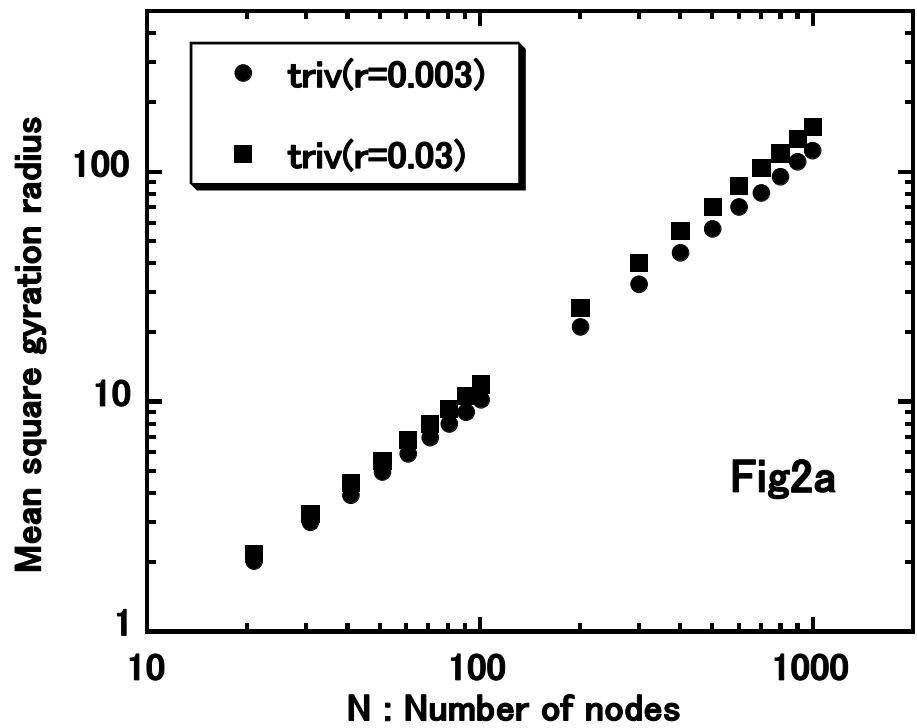
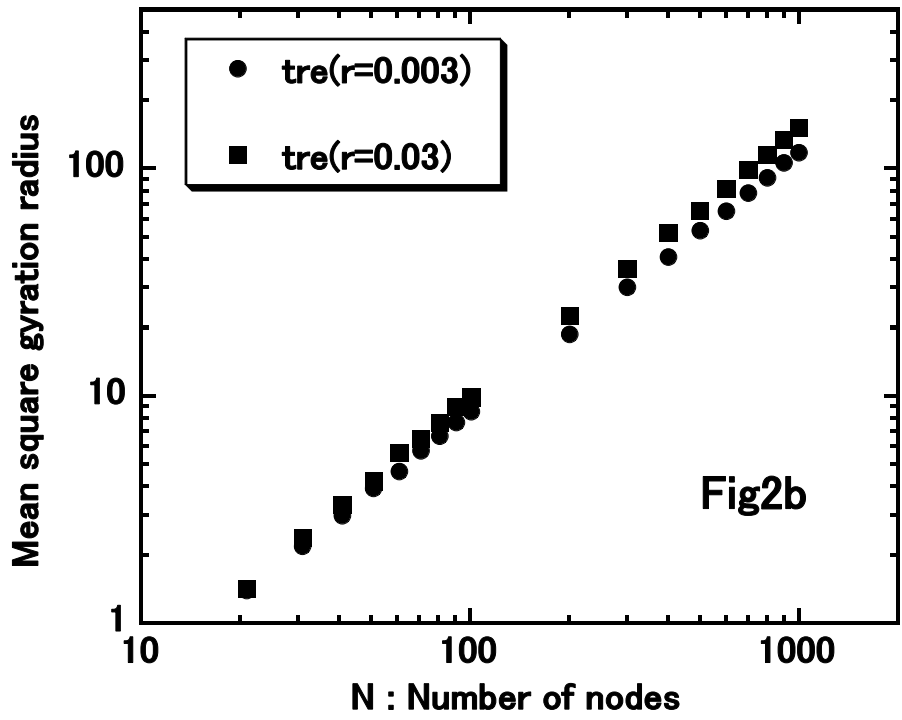
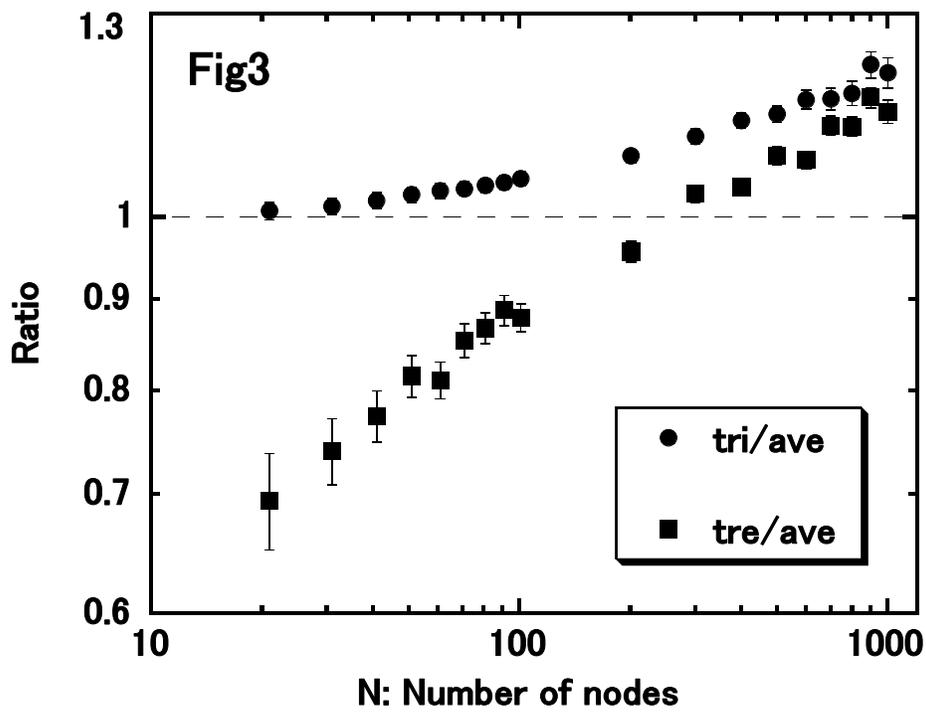
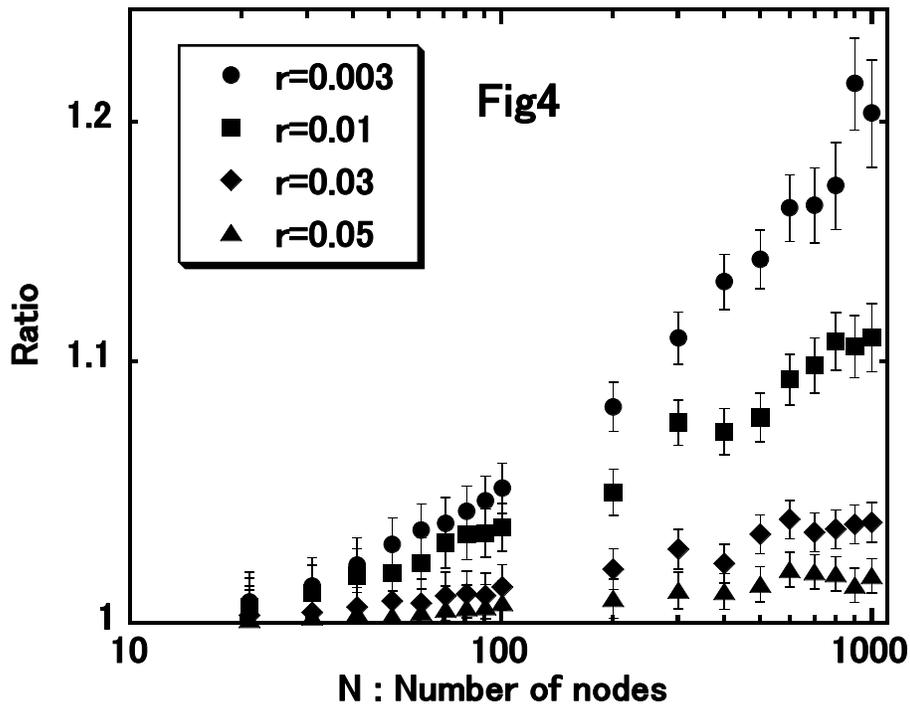
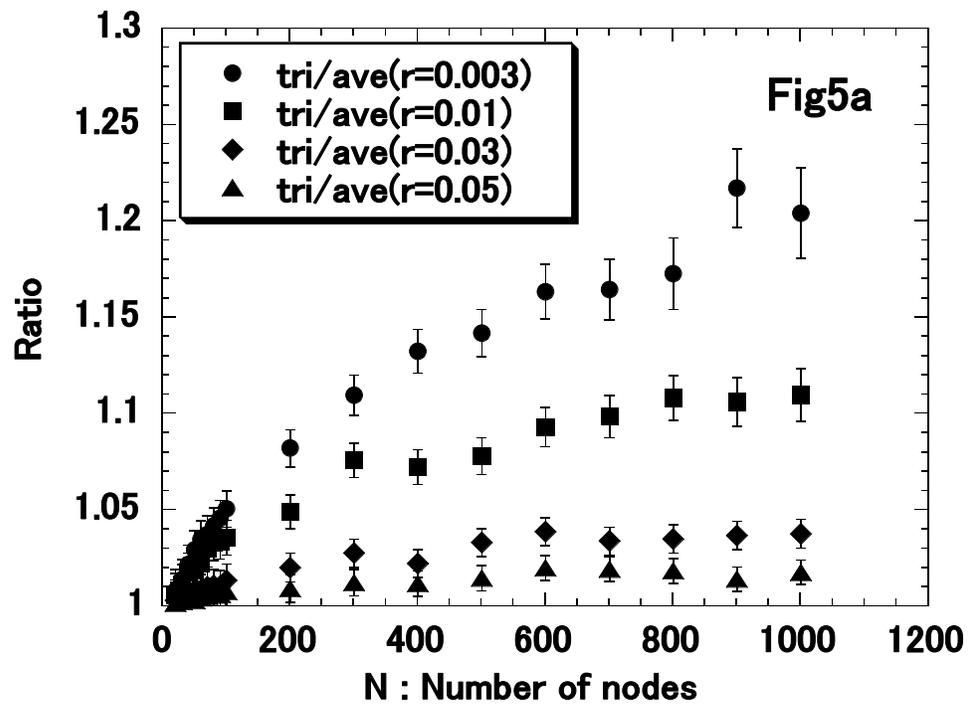
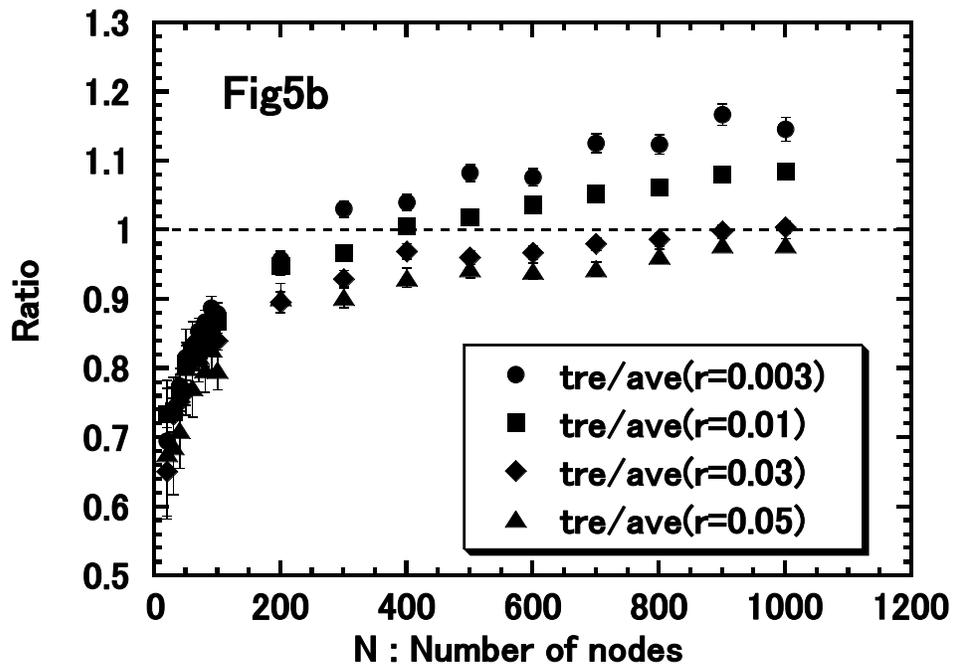
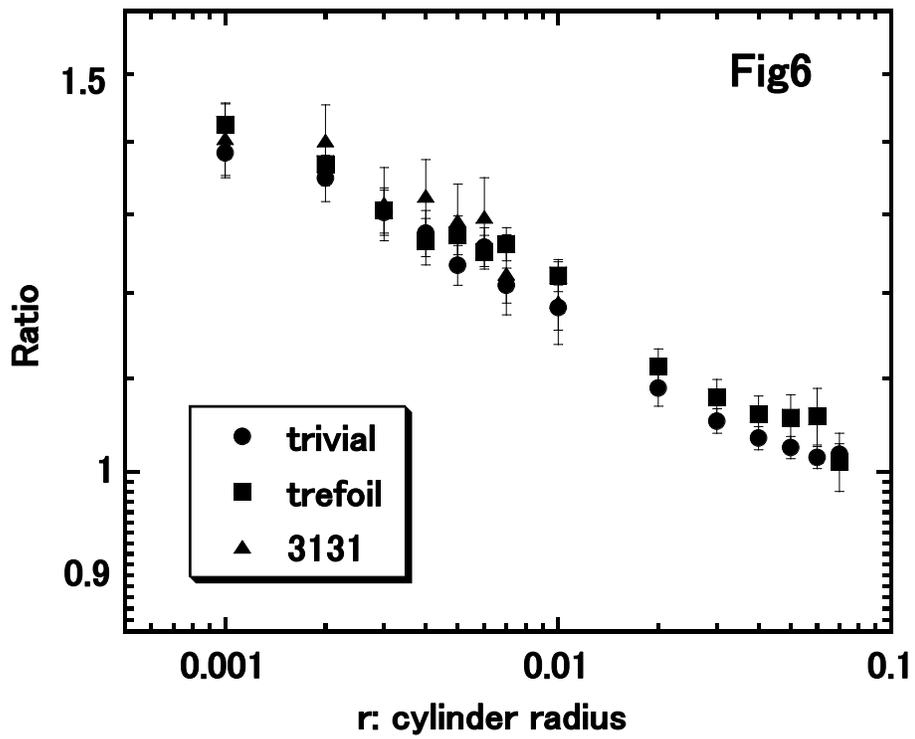
\end{document}